\documentclass[sigconf,9pt]{acmart}
\AtBeginDocument{%
  \providecommand\BibTeX{{%
    \normalfont B\kern-0.5em{\scshape i\kern-0.25em b}\kern-0.8em\TeX}}}

\setcopyright{none}
\acmDOI{}
\acmISBN{}

\acmConference[LIMITS '24]{Tenth Workshop on Computing within
Limits}{June
18--19, 2024}{Online}

\begin{document}

\title[Limits at a Distance]{Limits at a Distance: Design Directions to Address Psychological Distance in Policy Decisions Affecting Planetary Boundaries}

\author{Eshta Bhardwaj}
\email{eshta.bhardwaj@mail.utoronto.ca}
\orcid{}
\affiliation{%
  \institution{University of Toronto}
  \streetaddress{}
  \city{Toronto}
  \state{Ontario}
  \country{Canada}
  \postcode{}
}
\author{Han Qiao}
\email{h.qiao@mail.utoronto.ca}
\orcid{}
\affiliation{%
  \institution{University of Toronto}
  \streetaddress{}
  \city{Toronto}
  \state{Ontario}
  \country{Canada}
  \postcode{}
}
\author{Christoph Becker}
\email{christoph.becker@utoronto.ca}
\orcid{}
\affiliation{%
  \institution{University of Toronto}
  \streetaddress{}
  \city{Toronto}
  \state{Ontario}
  \country{Canada}
  \postcode{}
}

\renewcommand{\shortauthors}{Bhardwaj et al.}

\begin{abstract}

Policy decisions relevant to the environment rely on tools like dashboards, risk models, and prediction models to provide information and data visualizations that enable decision-makers to make trade-offs. The conventional paradigm of data visualization practices for policy and decision-making is to convey data in a supposedly neutral, objective manner for rational decision-makers. Feminist critique advocates for nuanced and reflexive approaches that take into account situated decision-makers and their affective relationships to data. This paper sheds light on a key cognitive aspect that impacts how decision-makers interpret  data. Because all outcomes from policies relevant to climate change occur at a distance, decision-makers experience so-called `psychological distance' to environmental decisions in terms of space, time, social identity, and hypotheticality. This profoundly impacts how they perceive and evaluate outcomes. Since policy decisions to achieve a safe planetary space are urgently needed for immediate transition and change, we need a design practice that takes into account how psychological distance affects cognition and decision-making. Our paper explores the role of alternative design approaches in developing visualizations used for climate policymaking. We conduct a literature review and synthesis which bridges psychological distance with speculative design and data visceralization by illustrating the value of affective design methods via examples from previous research. Through this work, we propose a novel premise for the communication and visualization of environmental data. Our paper lays out how future research on the impacts of alternative design approaches on psychological distance can make data used for policy decisions more tangible and visceral.

\end{abstract}


\maketitle

\section{Introduction} \label{introduction}

Decision-making for climate action and environmental protection relies on data communicated through descriptions and visualizations. When policymakers working to address large-scale issues like climate change, they rely on risk and prediction models, spatial analyses, and other climate services that act as decision-support tools for policymakers (for example, \cite{gutierrez_review_2019,kourtit_big_2018,pantalona_decision_2021,pignatelli_spatial_2023}. Trade-off decisions related to climate change have characteristics of risk, uncertainty, and ambiguity \cite{christoph_becker_searching_2023}. In our current times, the prioritization of economic growth means that sustainability is framed as a trade-off, and sacrificed \cite{escobar_encountering_2011,Hickel_2020}. As a result, humanity is breaching the safe planetary boundaries of our planet at an accelerating pace \cite{persson_outside_2022,richardson_earth_2023,rockstrom_planetary_2024}. Consider the following scenario: 

{\itshape Taylor is a transportation engineer at the City of Toronto’s Cycling and Pedestrian unit as part of the Transportation Services department. She has been asked to provide her opinion on whether a 10km  bikeway project should be approved in which a new bike lane would be constructed to connect existing bikeways in the Toronto and East York district. She reads a new report she received about active travel infrastructure that was implemented in boroughs in London, UK and how that led to changes in travel behaviour and health economic benefits that would continue to be seen for the next 20 years \cite{aldred_impacts_2024}. The report states that the cost of the program was £100 million \cite{aldred_impacts_2024}.}

While research has shown that active travel interventions provide up to 100 times the health economic benefits than the costs of their implementation \cite{aldred_impacts_2024,walker_health_2024}, the example above hints at the complexity of such a decision. The report Taylor reads to obtain information relevant to her decision is about London, with benefits accumulating over a 20-year time period but with upfront costs. It is unclear to her whether the benefits seen in London will apply in Toronto. Additionally, given that Taylor does not live in the Toronto and East York district, she has not yet and will not in the future experience the changes that a bike lane may introduce. 

{\itshape Jordan is a data specialist in the Data and Analytics unit of the Transportation Services department for the City of Toronto. She has been asked to develop and present a dashboard at a meeting next week for policymakers, sustainability experts, and engineers to review and decide whether a new bikeway project should be initiated in the Toronto and East York district. She gathers data about the geographic area in which the bike lane would be constructed, the projected costs and benefits, and presents it alongside other completed projects in Toronto as well as the figures presented in the London active transportation study. The dashboard is constructed using map-based visualizations, time series graphs, and KPI indicator visuals.}

% Psychology research provides insights into situations as the above.
% how to study decision-making which can be further leveraged to improve the design of tools used by policymakers to interpret and utilize environmental data. 
The social psychology concept of \textit{psychological distance} sheds light on situations as the above by explaining how decision-makers perceive and evaluate outcomes at a distance \cite{liberman_psychological_2007}. People face four dimensions of psychological distance while making decisions about a future scenario that is unlike their current circumstances due to the event occurring much later in the future (temporal), occurring at a geographic distance (spatial), perceived as not real (hypothetical), or considered to be outside their personal surroundings (social) \cite{fujita_psychology_2015,liberman_psychological_2007,soman_psychology_2005}. 

Given that policy decisions about climate change are urgently needed for immediate transition and change, designers of decision tools must recognize the psychological distance experienced by users when confronted with environmental data. Considering also that we need to anticipate drastic changes in a world that is increasingly destabilized, we will need a design practice that helps us with envisioning futures drastically different from the present.

In this paper, we explore the role of alternative design approaches in developing visualizations used for policy and decision-making given this context. We examine the following research questions: 
\begin{enumerate}
  \item How can research on psychological distance inform and shape data visualization design for policymakers working in the context of climate change?
  \item How can alternative design approaches support designers and decision-makers in envisioning radically different futures?
\end{enumerate}
    
We conducted a literature review and synthesis to bridge psychological distance research with design practices in computing to support climate policymaking. Based on our review, we claim that speculative design and data visceralization are promising avenues for conveying environmental data because they invoke affective (i.e., emotional) responses that stimulate policymakers to traverse psychological distance during decision-making processes. We propose four directions for future research to further the intersection between these fields and approaches. We contribute to the field of computing and especially the Limits community by surfacing the relationship between design practices and psychological distances and laying out potential research directions to further explore and evaluate this relationship for addressing climate issues. 

Below we first review concepts from social psychology and environmental psychology, then propose how speculative design and data visceralization approaches have been applied to make environmental data more tangible and visceral. We then develop key propositions that (should) form the core of future research directions and provide initial pathways that illustrate how these propositions can be explored in research projects. 

\section{Background} \label{background}
\subsection{Psychological Distance} \label{psychologicaldistance}

Psychological distance can be understood through construal level theory (CLT) borne out of social psychology. CLT explains the psychological processes that occur when travelling psychological distance \cite{liberman_psychological_2007}. Particularly, when making decisions, we naturally start at the current time and place, i.e. our reference point is egocentric. This means that any event that occurs close to this notion is considered proximal, while those farther are psychologically distant. This does {\itshape not} mean that they are automatically less important, but their distance influences how we perceive, represent, evaluate, and assess these events.

There are four dimensions to psychological distance (social, hypothetical, temporal, and spatial). All dimensions require similar psychological processes to traverse them \cite{fujita_psychology_2015}. CLT states that to traverse psychological distance, we construct representations of the distant event based on the knowledge we have at hand, also called creating construals \cite{trope_construal-level_2010}. People engage in cognitive abstraction for construal construction, which is a process in which the overall essence of a situation is extracted and details are abandoned. CLT proposes that when engaging in construal construction, we describe distant events using abstract features (high-level construal) that remain constant across time, space, etc. and proximal events using concrete situation-specific features (low-level construal) \cite{fujita_psychology_2015}. The relationship between psychological distance and level of construal has also been described as bidirectional \cite{rim_what_2013}. The effect of psychological distance on construal indicates that people use high-level construals to describe psychologically distant events because the essence of the events will not change with reduced distance (and the high level features will remain accurate while low-level features may change as details of the event become more tangible). The effect of construal on psychological distance means that an abstract, high-level construal will make the event feel more distant \cite{rim_what_2013,trope_construal-level_2010}. For example, an abstract concept such as ``having fun'', as compared to ``playing basketball'', invokes thoughts of distant activities along all dimensions of psychological distance \cite{trope_construal-level_2010}. 

The perception of climate change is also prone to psychological distance. For example, policymakers in Global North countries tend to face spatial distance from climate change as its impacts are experienced more intensely in other countries \cite{escobar_encountering_2011}. They also face temporal distance because the consequences of current decisions will be endured by future generations rather than occurring in our lifetimes \cite{gardiner_perfect_2011}. Policymakers also encounter social distance because vulnerable populations are affected first and to a greater extent \cite{farbotko_first_2012,hallegatte_climate_2017}. Lastly, they experience hypothetical distance because decisions made today about the future can have ambiguous and uncertain outcomes \cite{camerer_recent_1992}. Current decision-making for climate action employs data-driven visualizations including climate indicators, dashboards, risk modelling, etc. \cite{cepero_garcia_visualization_2020,gutierrez_review_2019,kourtit_big_2018,pantalona_decision_2021,pignatelli_spatial_2023,tikul_ptad_2022,wang_spatial_2017,zheng_stare_2022}. Therefore, the traversal of psychological distance has to occur through these tools while making decisions. 

{\itshape Taylor attends the meeting where Jordan presents the dashboards to their colleagues. Jordan first presents a feasibility analysis for the streets on which bike lane construction is being considered. It shows the speed thresholds and daily traffic volume for quiet versus fast, busy streets which informs the feasibility of the type of cycling facility that would be required (e.g., buffer, curb separator, bollard separator, signs and markings, etc.). Jordan further presents a projected impact analysis and a breakdown of costs that compares different cycling facility implementations. Although the presentation contains a variety of statistics and analyses relevant to the construction of bike lanes, Taylor remains unsure about the precise consequences of the decisions they have to make. As Taylor tries to visualize how the implementation of a dedicated bike lane would affect traffic flow in the area, there seem to be too many factors to arrive at an unambiguous outcome. For example, how will constructing a bike lane at a busy street with multiple retail businesses affect the number and severity of accidents? Which cycling facility is the most appropriate for each street section considering its busyness, the demographic of the area, and the projected traffic flow?}

People perceive climate change as distant unless they have personal experience that evoke emotional responses and visceral memories \cite{chu_risk_2020,weber_experience-based_2006}. Studies find that stating climate change communication through a risk-framing led to greater engagement, i.e., support for mitigation policies and mitigation intention, at greater spatial distance \cite{chu_risk_2020} and willingness to reduce energy use, at a greater social distance \cite{spence_psychological_2012}. Similarly, Weber argues that the ``...[affective system] has much greater influence over decisions under risk and uncertainty (including actions to address global warming)... Visceral reactions like fear or anxiety serve as early warning to indicate that some risk management action is in order and motivate us to execute that action'' \citep[p.~104]{weber_experience-based_2006}.

Since CLT argues that with increased psychological distance, construal shifts to higher levels, it could be predicted that the reverse is true, that enabling people to make lower-level construals would decrease the perception of distance and make climate change more concrete. However, studies show inconsistent results in the relationship between construal level and psychological distance in the context of climate change \cite{manning_psychological_2018,mcdonald_personal_2015,wang_climate_2019}. While studies have found that climate change is perceived as psychologically distant, it is unclear to which extent high versus low-level {\itshape construals} impact perceived psychological distance \cite{keller_systematic_2022, maiella_psychological_2020,wang_climate_2019}. An extensive review of research on psychological distance in the context of climate change shows a lack of consensus between studies due to differing measurement styles and reference frames \cite{keller_systematic_2022}.
%removed: Studies on the psychological distance of climate change also sometimes misapply CLT by measuring its direct impact on decisions rather than the information used to influence decisions \cite{keller_systematic_2022}. 

\subsection{Environmental Psychology } \label{environmentalpsychology}

The discipline of environmental psychology emerged in the 1960s and ``... includes theory, research, and practice aimed at improving human relations with the natural environment and making the built environment more human'' \citep[p.~543]{gifford_environmental_2014}. It considers both the impacts of humans on the environment and how the environment shapes human experiences \cite{steg_environmental_2018}. As an interdisciplinary field, it intersects with architecture and geography, social and cognitive psychology, and environmental science \cite{canter_environmental_1981,steg_environmental_2018}. The architecture and geography work focuses on built settings, design, and understanding how behaviour is impacted by physical-spatial surroundings. This particular aspect of environmental psychology was how the field was initially established and is why it was earlier titled as architectural psychology. As the field broadened, social and cognitive psychology influences led to the study of how the environment is perceived, spatial cognition, and proenvironmental behaviour \cite{gifford_dragons_2011,steg_environmental_2018}. At the same time, research on proenvironmental behaviour led to collaboration with environmental scientists \cite{steg_environmental_2018}.

As sustainability concerns became prominent, environmental psychology turned to the examination of proenvironmental behaviour in order to encourage human behaviour towards reducing negative environmental impacts \cite{steg_environmental_2018}. ``Pro-environmental behaviour refers to behaviour that harms the environment as little as possible, or even benefits the environment'' \citep[p.~309]{steg_encouraging_2009}, such as reducing air travel \cite{eriksson_necessity_2020}. Gifford describes key considerations as studying the influences and barriers to proenvironmental behaviour and their entanglement, which types of environmental behaviour occur in which social settings, and how social, political influences impact proenvironmental behaviour \cite{gifford_environmental_2014}. Given that we focus on the psychology of policymakers in this paper, we discuss the first challenge in more depth here from the perspective of how policymakers’ psychological processes can lead to their own proenvironmental behaviour (as compared to how they can promote this behaviour in the public). 

Proenvironmental behaviour is influenced by many factors. Our environmental concern and behaviour is influenced by childhood experiences (exposure to nature in childhood), knowledge and education about the environment, personality (openness, agreeableness, and conscientiousness), perceived behavioural control (i.e., self-efficacy), values, attitudes, and worldviews (altruistic, prosocial, biospheric values and ephemeral beliefs), felt responsibility and moral commitment (responsibility stemming from guilt), place attachment (strong emotional connection to a place), norms and habits (perceiving proenvironmental behaviour as the standard), affective reactions, and demographic factors (older, wealthier, and those living in rural areas) \cite{gifford_environmental_2014,gifford_personal_2014}. Particularly, studies urge that the impact of affective responses on proenvironmental behaviour should be taken into further consideration in climate policy \cite{brosch_emotions_2023,goldberg_emotion_2023,mishaud_affective_2023}. For example, Weber and Constantino highlight that IPCC reports have limited mentions of how emotions impact decision-making despite extensive research that demonstrates ``conceiving risks as `feelings' that people experience rather than as statistics better predicts their risk perceptions and financial actions'' \citep[p.~294]{weber_all_2023}. 

An extensive review of barriers to proenvironmental behaviour change \cite{gifford_dragons_2011} identifies 29 barriers and categorizes them into 7 groups. These include ``limited cognition'' (understood as limitations in human rationality), ``ideologies'' (beliefs and values that impede climate action, including ``technosalvation'' such as geoengineering), ``comparisons with others'' (perceiving others’ inaction as a reason for your own inaction), ``sunk costs'' (habitual behaviours, goals, or investment that resist change), ``discredence'' (mistrust and denial in climate change), ``perceived risks'' (perception of types of risks when considering a shift to proenvironmental actions), and ``limited behaviour'' (lack of positive feedback after climate action that limits further proenvironmental behaviour) \cite{gifford_dragons_2011,gifford_environmental_2014}. 

Specifically, research on the psychology of {\itshape climate change} considers climate change beliefs and values as central to promoting climate action. People are said to exhibit four kinds of values that guide their climate-related goals and motivations - hedonic (aim to improve pleasure and reduce effort), egoistic (aim to increase wealth and power), altruistic (aim to better societal welfare), and biospheric (aim to better environmental welfare) \cite{bouman_values_2021,steg_psychology_2023}. People with strong biospheric values demonstrate belief in climate change as real and are more likely to act on those beliefs. 

The value-belief-norm theory explores how 
values influence behaviour through beliefs and personal norms  \cite{stern_value-belief-norm_1999}. This means that people with greater biospheric {\itshape values} believe that their actions can have positive consequences which enables them to change or strengthen their personal {\itshape norms} in being more active in mitigating climate change \cite{stern_value-belief-norm_1999}. In those that have weak biospheric values, motivation towards greater climate action can be invoked by strengthening biospheric values including challenging and invoking reflection on their values, reinforcing prior climate actions, promoting environmental self-identity, perceiving oneself as part of a community, group, or organization that promotes and engages in environmental action, and making climate action the norm in society \cite{steg_psychology_2023}. On the other hand, stronger biospheric values also invoke more negative affective responses towards climate change (such as worry) \cite{bouman_when_2020} and negative affective responses towards climate inaction (such as guilt and shame) \cite{mandic_gen_2023}. Eco-anxiety and shame can lead to greater climate action and mitigation behaviour \cite{bouman_when_2020,mandic_gen_2023,martin_biospheric_2023}. 

\subsection{Disciplinary Framings} \label{disciplinaryframings}

Our review of past research here covers environmental psychology and social psychology and is influenced by behavioural economics, judgment and decision-making, and awareness of the tension between rationalist and naturalist perspectives on decision-making \cite{klein_sources_2017,Kahneman_Klein_2009}. Different disciplines frame observations of climate change decision-making in varying and sometimes conflicting ways. Selecting and adopting frames from other disciplines into computing must therefore be deliberate and thoughtful. 

Social psychology explains how social processes with other humans impact our thoughts and actions while cognitive psychology explains this through the lens of information processing such as the mental processes of perception and memory \cite{smith_social_2014}. Behavioral economics in turn leans on cognitive psychology but has adopted a particular framing of decision-making and rationality. While the broader field of judgment and decision-making includes the examination of both social and cognitive processes, it began with a normative orientation derived from economics and focused on behavioural theories of how decisions were made by supposedly ``rational'' individuals \cite{daniel_kahneman_prospect_1979,kahneman_judgment_1982}. A large segment of behavioral economics focused on measuring how human decision-making deviates from the mathematical models. These studies often theorized that in situations in which people did not make the ``rational'' choice, i.e. did not optimize utility effectively, it was because humans were  limited by their cognitive (information processing) abilities \cite{gigerenzer_how_1991,kahneman_subjective_1972,simon_behavioral_1955}. With further evolution in the field, debates between normative (how decisions should be made) and descriptive approaches (how decisions are actually made) became more prevalent and some focus shifted towards the impact of culture, context, and emotions on decision-making \cite{keren_birds-eye_2015,savani_culture_2015,fujita_psychology_2015}. However, utility models and the study of rational decision-makers are still key aspects of the disciplines of behavioural sciences and behavioural economics \cite{soman_psychology_2005}. The challenge with underlying rationalist assumptions is that they can limit our understanding of the full human reasoning capacity. ``By treating the cognitive process as machinery, the rationalist model prematurely abstracts the nuances'' of how people really make decisions \citep[p.~543]{fagerholm_its_2023}. The consequence is often to dismiss the capacity for humans to think long-term \cite{krznaric_good_2020} and treat human decision-making as a buggy algorithm that needs to be propped up and `fixed' by persuasive (or manipulative) design. Sustainable HCI research has long argued that such research is very limited in value and impact \cite{Bremer_Knowles_Friday_2022,Brynjarsdottir_Håkansson_Pierce_Baumer_DiSalvo_Sengers_2012}. In addition, the underlying assumptions distort how we study decision-making at all \cite{christoph_becker_people_2023}.  

Environmental psychology, with influences from social and cognitive psychology, remains largely consumer-focused. It still offers extensive research on what causes the assumed `general public' to demonstrate proenvironmental behaviour and how to increase this behaviour \cite{gifford_environmental_2014,steg_psychology_2023}. However, studies are often performed with participants from unrepresentative populations like North American undergraduate students, which skews the findings \cite{henrich_weirdest_2010,thalmayer_neglected_2021}. Similarly, behavioral economics also has a consumer orientation. Within that context, social psychology research from which construal level theory and psychological distance emerge proposes that psychological distance can be reduced by generating low-level construals \cite{liberman_psychological_2007,trope_construal-level_2010}. Much of this work continues to focus on consumer choices, but is also applied to the much larger question of climate change, where its findings are debated \cite{keller_systematic_2022}. The inherited framing of decision-makers as consumers can impose severe limits on what can be studied and how we should interpret the findings of these studies. Beyond being exclusively individual participants of a fragmented `public' consumer society in which they make nothing but consumer choices, humans also act collectively and form {\itshape publics} to debate how to change the system \cite{Bhardwaj_Qiao_Becker_2023,Dantec_2016}. 

Our position on these disciplinary framings is that we must be cognizant of their limitations when drawing on research from these fields. It is worth noting however that CLT assumes neither a consumer orientation nor a rationalist framework.  

\subsection{The Psychological Dimension of the Limits Research Agenda} \label{limitsresearchagenda}

The trouble with {\itshape Limits} is that they always seem far away. We need to bring the far-away into view to act on it. That our societies’ failures to do so leave us ``deviant and guilt-ridden'' has been discussed since the founding of this research community \cite{Knowles_Eriksson_2015,eriksson_exploring_2023}. 

If, as many suggest \cite{bill_tomlinson_introduction_2012}, computing is to be a tool to help us cope with the massive scope in time, scale, and complexity on which the limits of the safe planetary space turn from a ``risk'' concept into a material reality, then the psychological distance frame is crucial for the Limits research agenda. Computing and information technology in general enable people to perceive situations that extend to scales beyond a sense of self \cite{bill_tomlinson_introduction_2012}. For example, it broadens time and space by making long-term and global collaborations possible and can alleviate social and disciplinary complexity \cite{bill_tomlinson_introduction_2012}. Similarly, designing decision tools to aid in the traversal of psychological distance can be a way to compress time, space, and complexity. 

To orient Limits research at this challenging juncture, we need to overcome the limitations of dominant but flawed paradigms and pay attention to the cognitive, psychological, and social dimensions of shifting computing into the safe planetary operating space. Below we demonstrate this orientation in exploring how we could design systems that help policymakers to traverse the psychological distance that separates them from the outcomes of their policies.

\section{Addressing Psychological Distance through Design} \label{addresspsychdist}
In this section, we explore how the design and development of decision tools can aid in the traversal of psychological distance by policymakers. Specifically, we synthesize literature from environmental psychology that explains the influence of affective reactions on proenvironmental behaviour with social and cognitive psychology which discusses that reducing psychological distance to climate change invokes climate action. Further information about our method for the literature review can be found in Appendix \ref{ap_litreview}. We propose that decision tools can leverage speculative design approaches and data visceralization to invoke affective responses to climate change in order to traverse psychological distance. We introduce these methods and present illustrations of their application below. 
\subsection{Speculative Design} \label{speculativedesign}
Critically oriented design practices in computing offer views on the future of computing that are drastically different from ideas such as solutionism and growth, some of the underlying assumptions that the field has grown out of \cite{sharma_post-growth_2024,wong_speculative_2018}. Speculative design surfaces values, critiques social issues, and presents alternative visions of the future by creating conceptual proposals and artifacts \cite{khovanskaya_double_2015,wong_infrastructural_2020}. Given that climate change is a complex problem shaped by many underlying assumptions about social, economic, and political aspects of human lives, we see potential in adopting speculative design to bridge psychological distance. Through such design practices, we will have unique opportunities to radically reimagine worlds that we want to live in and re-evaluate the potential as well as the limits of computing in working towards these futures. In this section, we review the speculative design approach and its benefits and critiques, and then highlight its possible application to climate change policymaking.

The term ``speculative design'' originated from Dunne and Raby as a means of contemplating, discussing, and debating alternative futures through the lens of design \cite{dunne_speculative_2013,soden_what_2021,dunne_hertzian_2005,khovanskaya_double_2015,soden_what_2021}. The design outcomes can take form in various ways including design proposals, exhibits, design fictions, working prototypes and products etc., but they are never intended to be the solutions to a problem. The value of speculative design lies in the critical contemplation, discussions, questions, and debates about the current state of the world as well as possible futures that arise through experiencing the processes of designing or interacting with the design outcomes \cite{dunne_speculative_2013,wong_speculative_2018}. Past adoptions of speculative design practices can be found in both academia and industry \cite{wong_speculative_2018,wong_when_2016}. Research projects use speculative design to explore a wide variety of topics including smart homes \cite{aipperspach_heterogeneous_2008}, cameras \cite{pierce_variations_2014}, food systems \cite{heitlinger_co-creating_2019}, climate change \cite{soden_what_2021}, and their socio-political implications. Industry uses speculative design both as a means of internal vision setting \cite{kuester_speculative_2019} and as channels of communication to external stakeholders on research initiatives that transcend technical and commercial constraints \cite{harrison_methods_2001,wong_speculative_2018}. Speculative design is useful and charismatic for engaging a wide range of audiences in thought provoking experiences, but scholars have also noted its limitations. For instance, Oliviera, Prado, and Tonkinwise pointed out that speculative design usually lacks discussions of race and class \cite{prado_questioning_2017, tonkinwise_how_2014}. In the context of climate change, Anderson questioned whether speculative design could convey the urgency required for climate actions \cite{anderson_ethics_2015} and some commentators have questioned whether it trivializes serious issues
\cite{thackara_republic_2013}.

Acknowledging these pushbacks and concerns, design and computing scholars have argued that the benefit of speculative design lies in attempting to better understand existing problems, create new knowledge, and engage a much broader audience \cite{thackara_republic_2013}. Instead of presenting fully investigated truths of real-world issues, speculative design offers new angles to the increasingly complex social issues we face today by generating understandings that help inform future directions \cite{thackara_republic_2013}. As critics of speculative design, Oliviera and Prado also acknowledged that ``design is a powerful language'' and that ``envisioning near future scenarios might help us reflect on the paths we want to take as a society'' \citep[p.~5]{prado_questioning_2017}. 

When applied to climate change, Soden et al. suggested that speculative design brings four benefits: 1) allowing discussions around hopeful and optimistic futures, 2) questioning values and ethics embedded in current practices, 3) encouraging public discourse through participatory activities and 4) allowing personal narratives about the environment and the future \cite{soden_what_2021}. We hypothesize another benefit of applying speculative design in the context of climate issues:  the effect of bridging psychological distance and supporting civic engagement efforts in policymaking. The  paragraphs below surface how social, hypothetical, temporal, and spatial psychological distances are being traversed and bridged through past speculative design projects in the context of climate and computing.

\textit{Examples:} High Water Pants are a speculative wearable technology by Biggs and Desjardins that works by mechanically shortening when a cyclist enters an area that will be impacted by sea-level rise in 30-80 years \cite{biggs_high_2020}. The High Water Pants stimulate scenarios of future sea-level rise through using material artifacts for embodied speculation. The design creates opportunities for cyclists to speculate on the basis of their personal embodied experiences and place-based knowledge about their bike routes. In this project, cyclists are able to bridge temporal distance, or in the authors’ words time is ``bent'' \citep[p.~9]{biggs_high_2020}, because they are allowed to presently feel the future of our world and speculate about the ramifications of that future. Spatial and hypothetical distance are also addressed through this design as in the authors’ exploratory interview with cyclists, they identified participants’ ``difficulty locating [climate change] spatially'' \citep[p.~6]{biggs_high_2020} and even when participants mentioned that they were sad about climate change, they experienced difficulties ``pinpointing tangible evidence of climate change in their daily commute.'' \citep[p.~6]{biggs_high_2020}. Through speculative design, the High Water Pants were able to open a space for local cyclists to notice potential future impacts of climate change by allowing them to experience a possible future at specific geographic locations based on existing hypotheses of climate change. In the authors’ words, speculative design can ``bend'' larger scales to ``a scale that humans can feel, or at least, imagine feeling'' \citep[p.~9]{biggs_high_2020}. The nature of speculative design in inviting someone to be somewhere, at some time helps participants traverse different kinds of interrelated dimensions of psychological distance.

Another speculative design project led by Heitlinger et al. engaged local urban growers and local small business stakeholders to explore sustainable food futures \cite{heitlinger_co-creating_2019}. Four participatory speculative design workshops included mapping future gardens, neighborhood walks for imagining where and how food could be grown in the future city, games of envisioning future food growing in speculative lands, as well as worldbuilding using crafts. Participants were able to traverse multiple dimensions of psychological distance through these speculative workshops. Firstly, social distance between citizens, community urban growers, and small business stakeholders was bridged through the participatory nature of speculative design as they engaged in these workshops together to collectively reimagine future urban food-growing beyond corporate visions. Spatial distance was traversed through the neighborhood walks in the second workshop, as participants were able to see, smell, hear and experience activities  in the specific areas of their neighbourhoods and engage  with ``space-specific stories, metaphors and materials used for futuring.'' \citep[p.~5]{heitlinger_co-creating_2019}. In the third workshop, through engaging in a board game, the participants were brought into speculation exercises where their lands were either developed into utopian or dystopian future scenarios and they discussed visions of future food growing within these lands, bridging hypothetical distance.

These examples of past research bridge psychological distance by leveraging speculative design to open up design spaces, provoke reflection, and facilitate long term pathways to address climate issues. Although speculative design’s relationship with social psychology was not explicitly discussed in the original research, we see potential in connecting the two concepts to enrich both fields by contributing new knowledge that informs future designers of computing tools and policymakers who hope to tackle climate change issues. Speculative design offers an alternative vision of our world, supporting us to traverse to a time in the future, grounding us to a specific location, and anchoring us to a concrete hypothesis of the future. Through traversing psychological distance, we are encouraged and empowered to radically imagine worlds beyond the one assumed today, dominated by capitalism and limitless expansion.

\subsection{Data Visceralization} \label{datavisceralization}

The field of information visualization has long followed the paradigm that objective, neutral, and minimalistic visualizations represent the gold standard. Tufte’s concepts on chart junk, data-ink ratio, and achieving graphical excellence and integrity have dominated as best practices for developing visualizations \cite{tufte_visual_2013}. However, current research has established that data representations are never neutral and there is no utility in attempting to be 
\cite{dignazio_rational_2020,li_data_2018,van_koningsbruggen_its_2021}. D’Ignazio and Klein underscore that all visualizations, no matter how minimal, are constructed by designers who make choices that result in framing effects on how they are interpreted \cite{dignazio_rational_2020}. With this perspective, no data visualization is neutral, and visceralization becomes another framing through which emotional responses can be elicited. It becomes a tool that can be leveraged to convey data. For example, Schmidt et al. present a project in which a machine learning model is used to generate before and after images of places experiencing extreme weather events in areas that would not normally have these occurrences to invoke visceral reactions that create an improved understanding of climate projections \cite{schmidt_visualizing_2019}.

Prevailing methods of developing visualizations for dashboards and other tools are not sufficiently considering that people face psychological distance to climate change unless they have experienced its impact personally which causes emotional responses and memories \cite{chu_risk_2020,weber_experience-based_2006}. Data visceralization is aimed at eliciting affective responses that transcend a visual experience \cite{dignazio_creative_2017,dignazio_rational_2020}. This newly emerging approach to data visualization \cite{huang_walking_2021,jang_visible_2022,lee_data_2021} and story-telling \cite{armeni_narratives_2021,finnegan_its_2023,harris_telling_2017,zhou_data-driven_2023} has cognitive benefits in eliciting action by invoking affective reactions to perceived risk \cite{mcdonald_personal_2015,weber_experience-based_2006}. Our hypothesis is that making data visceral may support decision-makers in traversing psychological distance by making climate change more personal, less abstract, and closer in time and space. 

The visceralization of data can take many forms and is not limited to techniques of visualization. For example, Li proposes that creative art can present visceralized data to convey an experience through an aesthetic approach \cite{li_data_2018}. Similarly, von Ompteda discusses the use of art installations to invoke sensory experiences that reduce the psychological distance to climate change \cite{von_ompteda_data_2019}. Risen and Critcher’s study attempts to make impacts of climate change more localized and therefore visceral by influencing bodily states \cite{risen_visceral_2011}. They found that people feeling warm (due to an increased indoor temperature) believed in global warming, i.e., ``it was easier for participants to conceptualize and simulate the global warming emergency while sitting outside on an uncomfortably warm day than while huddling for warmth in a nippy wind'' \citep[p.~781]{risen_visceral_2011}. 

A recent paper by Lan et al. extensively analyzes how affective visualization is discussed by researchers including the reasons for which visualizations are created with affect in mind, what types of problems affective visualizations can be applied to, what form the visualizations can take, and how to design them \cite{lan_affective_2024}. Lan et al. find that the categorization of visualizations as ``affective'' occurs due to three reasons: (1) collecting data to relay emotions such as heart rate, (2) stimulating people to evoke an emotional response and evaluate whether this causes a change in perception, and (3) studies where visualizations are shown to people to measure whether they cause an emotional reaction \cite{lan_affective_2024}. For example, Risen and Critcher’s study of the connection between felt temperature and climate change beliefs \cite{risen_visceral_2011} falls into the second category whereas Li \cite{li_data_2018} and van Ompteda’s \cite{von_ompteda_data_2019} studies fall into the third category. Furthermore, Lan et al. found that researchers explained in their studies that affective visualization design (category 3) was important because it aided in data comprehension and made data more humane by making the people in the data more real. \cite{lan_affective_2024}.

%removing: for the following 5 reasons  They describe that certain disciplines value affective design, others considered affective design as useful for data comprehension, some studies echoed that there was no visualizations are neutral and therefore visceralization could be used as a rhetorical device, some further discussed that creating non-affective visualizations is a social construction that should not be heeded, and that visualizations should be more humane and consider emotions, particularly when related to real people \cite{lan_affective_2024}. %
%removing: %A further analysis of 61 affective visualization design projects by Lan et al. breaks down the disciplines in which they are applied (for example, environmental sciences and ecology, health and well-being, etc.), what the purpose of the visualizations is (for example, provocation, advocacy, etc.), and how they are designed (for example, narratives, artifacts, immersive environments, etc.) \cite{lan_affective_2024}.

\textit{Examples:} To appreciate the impact of affective visualizations, we discuss two specific examples in detail below. Studies that use virtual reality applications to narratively depict data define data visceralization as ``... a data-driven experience which evokes visceral feelings within a user to facilitate intuitive understanding of physical measurements and quantities'' \citep[p.~1095]{lee_data_2021} where viscerality means a ``subjective sensation of being there in a scene depicted by a medium, usually virtual in nature'' \citep[p.~1095]{lee_data_2021}. In the first example, virtual reality technology is used to provide an immersive experience of a simulated forest 50 years in the future using publicly available data \cite{huang_walking_2021}. The VR experience enabled users to explore two climate change scenarios and how different plant species respond to climate change \cite{huang_walking_2021}. Users reported on their ``sense of presence'' by answering questions about whether they felt physically present in the simulated forest, whether they felt they could move and interact in the environment, whether they felt their location had been shifted, etc. \cite{huang_walking_2021}. The high-quality VR simulation enabled users to have an immersive and embodied experience. While the VR experience may have transported users into a future forest, the reduction of the spatial psychological distance may not be visceral enough on its own. We can argue that there are degrees to viscerality. For example, in a study where participants were exposed to either a magazine article, a video, or virtual reality application of a wildfire (the climate change impact was previously established), results showed that the participants reported more intense emotions in the virtual experience \cite{meijers_experiencing_2023}. The authors reflected that the qualitative results demonstrated that the VR experience did reduce psychological distance however this may not have been captured in the quantitative results due to measurement issues~\cite{meijers_experiencing_2023}. 

Data visceralization can also be performed through visual installations in public areas for the purpose of education and advocacy, such as in the case of Kuznetsov et al.’s spectacle computing project on air quality \cite{kuznetsov_red_2011}. Balloons installed with air quality sensors that changed colour based on detection of exhaust gas, diesel, or volatile organic compounds were installed in a public park and city street in Pittsburgh, Pennsylvania, USA (rated as having poor urban air quality) \cite{kuznetsov_red_2011}. Observers initially mistook the balloons as signifying a nearby celebration but on closer look expressed concern about air quality, especially those who lived in the neighbourhood \cite{kuznetsov_red_2011}. In terms of gathering public attention, the balloon installation was a successful example of how ``playful media'', instead of persuading proenvironmental behaviour, can act as an entry point to climate education, discourse, and exploration \cite{kuznetsov_red_2011}. 

The above studies demonstrate that the use of visceralization and speculative design through affective data representation can be applied to climate change communication and visualization to reduce psychological distance. In the following section, we propose research directions for further examination of how these techniques can be applied to traverse the psychological distance experienced by policymakers when making environmental decisions. 
\section{Research Directions} \label{researchdirections}

To advance research on psychological distance and its impact on policymaking for environmental decisions, we explore research directions in the form of four propositions below. The first synthesizes a claim established by social and cognitive psychology research and proposes open questions. The second and third present research opportunities for studying the impacts of visceralization and the application of speculative design on psychological distance related to environmental decisions, respectively. The last claims that speculative design and data visceralization are complementary methods that can be used in studying psychological distance. 

\subsection{Evaluate Psychological Distance} \label{researchdir_prop1}

Our first proposition is that psychological distance can be evaluated. Construal level theory states that ``...people engage similar psychological operations to travel mentally across each of [the distance dimensions]'' \citep[p.~405]{fujita_psychology_2015}. There is extensive literature that evaluates the perceived abstraction of psychologically proximal or distant events using CLT, for example in consumer research and psychology \cite{huang_effects_2016,thomas_psychological_2012}, retail \cite{darke_feeling_2016}, media psychology of online and digital communication \cite{norman_distance_2016,sungur_psychological_2019}, health and disease communication \cite{kim_how_2019} to name a few as well as a meta-analysis of the effect of psychological distance on abstraction \cite{soderberg_effects_2015}.

Studies have also evaluated the traversal of psychological distance related to climate change. For example, Duan et al.’s experimental study showed 402 participants abstract versus tangible climate change images to study whether they invoked higher versus lower-level construals \cite{duan_abstract_2019}. Their findings demonstrate implications for visual communication to the public since participants who were shown more concrete images reported construction of lower-level construals \cite{duan_abstract_2019}. Similarly, Singh et al. conducted a survey of 653 participants in the US to understand how perception of climate change (by measuring self-reported spatial, social, and hypothetical distance) impacted support for climate adaptation and concern for climate change using mediation conceptual models \cite{singh_perceived_2017}. Jones et al.’s study of 333 Australian participants studied how framings of psychological distance (all dimensions) affected mitigation intentions using principal components analysis and path analysis \cite{jones_future_2017}. The diversity of the methods used in these studies as well as the research questions shows that the evaluation of psychological distance is a substantive area of research. 

The interaction between the dimensions of psychological distance has been established and examined \cite{fiedler_relations_2012,maglio_distance_2013,trope_construal-level_2010}, but some argue there is an asymmetrical impact on judgment \cite{zhang_psychological_2009}. Keller et al. argue the dimensions are traversed using different psychological processes and therefore should be studied separately, and with more diverse populations given the high influence of socio-demographic factors on experience of psychological distance \cite{keller_systematic_2022}.

\subsection{Visceralization can Reduce Psychological Distance} \label{researchdir_prop2}

To extend the first proposition, we recommend further research to examine how visceralization can help policymakers traverse the psychological distance to climate change and other distant environmental decisions. We present an illustration of a study that explores how data around active transportation interventions can be visceralized and how the impacts of visceralization can be evaluated. 

Staying within planetary boundaries requires proenvironmental behaviour. For consumers, this can mean changing their behaviour to adopt biking as their means of transportation rather than driving. For policymakers, proenvironmental behaviour would be to support the development of infrastructure that makes it possible for people to bike. The positive outcomes of this would lead to better health effects and better environmental impacts. In the North American context, one of the barriers to the uptake of biking is that it is dangerous because of the prevalence and severity of motorist-cyclist accidents due to lacking infrastructure. As such, it is the responsibility of policymakers to reduce these barriers by making biking a safe option. 

Our illustrative example study concerns policymakers exploring decisions about constructing a new bike lane. The study could be conducted with a participatory approach in which the inclusion and participation of policymakers would center their voices and first-hand experiences. By focusing on their expressed challenges and barriers to making decisions at a distance, the study would engage directly in understanding and overcoming their key concerns. From our earlier scenario, Taylor expressed difficulties in understanding the consequences of bike lane construction and articulating why a new policy is needed. The presentation of aggregate statistics result in abstract high-level construals. The associated ambiguity  creates an obstacle to action. Low-level construals can help overcome this inertia through concrete incidents with incidental detail. By visceralizing data, we can prompt lower-level construals and tap into the affective power of emotional reactions. That could help bring the  policymakers closer to the impact they have and makes consequences tangible and therefore less ambiguous. In other words, to evaluate whether a policy should be passed to construct bike lanes, policymakers can use visceral data presentations. 

Conventional time-series graphs can represent previous motorist-cyclist accidents as non-visceral. For visceral visualizations, classic spatial analysis can be augmented with experiential imagery to transport the data-user to the scene of previous incidents. The collisions can then be demonstrated viscerally by using a dynamic map-based plot with incidents as bike outlines on a street-view of the intersection on which they occurred. Furthermore, adding details such as the date of the incident and a description of the incident would make the incident more tangible. 

The effects of visceral visualizations on psychological distance experienced by policymakers can be evaluated using techniques adopted from CLT. For example, policymakers can be asked to self-report their perceived psychological distance and mental construal when shown non-visceral (such as dashboard shown by Jordan) and visceral (as described above) visualizations of motorist-cyclist accident data. The perceived psychological distance before and after seeing non-visceral versus visceral visualizations can be measured through self-reporting by asking policymakers to make considerations such as, ``motorist-cyclist accidents are likely to occur in my local area'' (geographical), ``I am likely to be involved in a motorist-cyclist accident'' (social), ``this data is representative of current motorist-cyclist accidents'' (temporal), ``I am uncertain that bike lanes will impact the number of motorist-cyclist accidents'' (hypothetical). This will result in a quantitative assessment of the effects of data viscerality on the perceived abstraction of bike lane construction causing psychological distance. These considerations could then contribute to the larger discussion and decision around bike lane construction. 

%%Note that visceralization can also be performed to elicit positive emotions.%%

% In our current example, we discuss the visceralization of motorist-cyclist accidents by evoking negative emotions. However, visceralization can also be performed by eliciting positive emotions. For example, a lack of bike lanes leads to less active means of transportation that rely on motorized vehicles. This can greatly increase pollution in a given area. To visceralize the beneficial health impacts of constructing bike lanes, policymakers can be shown a video in which travellers in an area with bike lanes have fresh air with smells of freshly cut grass, flowers, and summer rain to demonstrate the cleanliness of the air.  

We should evaluate: under which conditions do visceral visualizations impact policymakers’ ability to traverse psychological distance, and how? This can be studied by using knowledge elicitation methods from applied cognitive task analysis \cite{crandall_working_2006,klein_sources_2017,klein_critical_1989,klein_guidelines_2001,fagerholm_its_2023}. 
% This is a knowledge elicitation method (used to understand people’s knowledge, skills, and judgments) from CTA (a toolbox to study real-life cognitive processes) which has been widely adopted within software engineering, human-computer interaction, and other computing disciplines \cite{fagerholm_temporal_2019,fagerholm_its_2023,gutzwiller_task_2016,horn_adapting_2017,nasser-dine_does_2021,roose_tracer_2018,rosli_user_2016}. Using CTA will provide the framework to examine how individual dimensions of psychological distance are affected by data visceralization. 

\subsection{Speculative Design Helps Us Envision Alternative Futures} \label{researchdir_prop3}

Speculative design helps us in envisioning an alternative future that is radically different from the one today. In typical design workshops, most design assumptions are continuous from what we are doing now, most design goals are contributing to a world that is more or less the same as the one we inhabit now, and the design outcomes are stabilizing existing systems. Speculative design offers a discontinuity that results from humanity breaking the boundaries, or limits, of  existing systems where either collapse has advanced further or a safe planetary space has been re-established. Design for a discontinuous future is much needed in times of the climate crisis \cite{easterbrook_discontinuous_2021,knowles_this_2018,tomlinson_collapse_2013}, and speculative design is uniquely able to invite participants to these drastically different lifeworlds \cite{wong_infrastructural_2020}.

To extend the sample study discussed about supporting policymakers around the construction of bike lanes, speculative design workshops can engage residents and public servants to come together to imagine future cities. The workshop formats can be games, walks, crafts building, or bike trips to bridge different dimensions of psychological distance. One potential lifeworld for speculation could be where cars are banned from entering a city and urban lives would rely primarily on biking, walking, and public transport. Such speculation can take advantage of events in cities that temporarily create conditions that align with the speculation. For instance, the City of Toronto's High Park Movement Strategy transformed parts of the roads around High Park to be car-free to improve safety, accessibility and the park's natural environment \cite{toronto_high_2022}. A drastically different lifeworld would be entirely car dependent – an urban space full of big box stores, parking lots, and highways. Participants can then both individually and collectively imagine how they would want to live their lives in these dramatically different lifeworlds and come up with their own imagined futures.

Due to the complexity of issues related to climate change, policymaking often seems to be very constrained with social, economic, and political limitations. However, such complex and urgent issues require immediate, radical transformations, not gradual and incremental changes \cite{easterbrook_discontinuous_2021,klein_this_2014,Hickel_2020,knowles_this_2018}. To achieve this, speculative design offers a unique opportunity for engaging collective radical reimaginings of the possible future worlds beyond the limits of the existing norms. How can speculative design be applied to policymaking beyond its conventional context in designing artifacts? How can the participatory nature of speculative design be centered to avoid pseudo-participation \cite{palacin_design_2020} and instead allow diverse inclusive imaginations of futures \cite{escobar_designs_2018}?

% \vspace{-0.5cm}

\subsection{Speculative Design and Data Visceralization Complement Each Other} \label{researchdir_prop4}

% Under the context of policymaking for climate change, we see a potential application of speculative design in data visceralization and, reversely, application of data visceralization in speculative design. 
The two methods complement each other especially for bridging psychological distance. In the example of High Water Pants \cite{biggs_high_2020} discussed in the previous section, this two way relationship is highlighted: through speculative design, the pants were created to present sea-level rise data in a visceral way, and through wearing these pants and interacting with the visceral representations of sea-level data, cyclists were engaged to speculate how climate change impacted futures they hoped to live in.

\begin{figure}
    \centering
    \includegraphics[width=7cm,height=8cm]{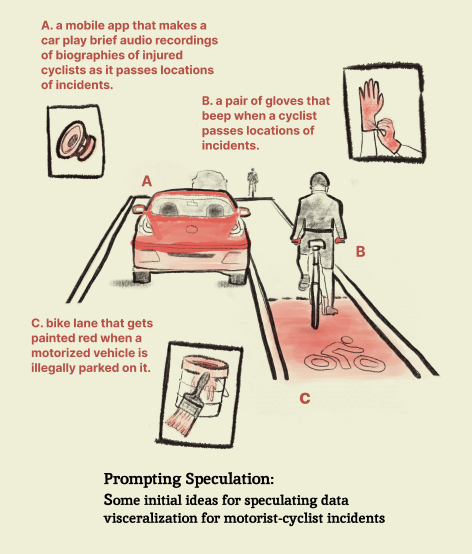}
    \caption{Speculative design ideations for visceralizing motorist-cyclist incidents data. The purpose of this visual is to illustrate potential ideas of visceralization, inspire discussions and debates but not to provide solutions.}
    \label{fig:fig1}
    \vspace{-0.2cm}
\end{figure}

%%We propose a research opportunity through the example of policymaking in constructing bike lanes.%%
Beyond the visceralization of augmenting data representation with experiential imagery discussed in Section \ref{researchdir_prop2}, future research could design and implement speculative design workshops to engage more local residents and policymakers to speculate other ways of data visceralization that prioritize local communities’ values. Through running a sample design exercise amongst the authors of this paper, we came up with a variety of ways of visceralization: an app that makes the car play brief audio recordings of biographies of injured cyclists as the phone passes locations of incidents, a pair of gloves that beep when a cyclist passes locations of incidents, and bike lanes that get painted red when a motorized vehicle is illegally parked in them (Figure \ref{fig:fig1}). The visceralization impacts perceived psychological distance because the individualized data (as compared to aggregate) and application of the identifiable victim effect (increased empathy for identifiable people as compared to collectives) \cite{fujita_psychology_2015} reduce social distance, showcasing the exact location of the accident reduces spatial distance, and the accidents are seen as less distant in the hypothetical dimension because they have already occurred. These designs might not be good or workable designs, or ones that provide direct solutions, but they are helpful in raising questions around who bears the burden of traffic incidents and how we can empower vulnerable road users through road design. Note that visceralization can also be performed to elicit positive emotions. A design workshop that involves residents and policymakers would generate more valuable and meaningful ways of visualizing data. These outcomes of speculative design will support stakeholders involved in policymaking to speculate a future that they hope to live in that centers justice and human dignity.

Although this section only presents one possible application of combining speculative design with data visceralization, we see value in exploring how speculative design can be applied to and combined with data visceralization in other ways. We invite future research to explore how the two practices can be combined.
% through applying them to real life cases in this age of rapid datafication and propose alternative paradigms for collecting, using, and communicating climate related data and their relationships with policymaking and design.

\section{Discussion} \label{discussion}

\subsection{Psychological Distance and Climate Action} \label{psychdist_climact}

Above we discussed how the reduction of psychological distance is considered to make impacts of climate change more proximal, thereby increasing the likelihood of a person to act on it because it is more personally relevant. However, many studies demonstrate cases in which no such effect is observed \cite{chu_taking_2018,maiella_psychological_2020,van_valkengoed_psychological_2023,wang_climate_2019}, i.e., ``seeing climate change as more distant does not necessarily result in less climate action, and reducing [psychological distance] does not reliably increase climate action'' \citep[p.~362]{van_valkengoed_psychological_2023}. Brügger et al. assert that ``...varying levels of psychological distance (e.g., here vs. far away) influence {\itshape how} people represent objects mentally and {\itshape what} information they consider when making judgments and decisions.'' \citep[p.~125]{brugger_proximising_2016}. In accordance with this perspective ``...proximising climate change should affect {\itshape how} climate change is mentally represented, and through this what people {\itshape act on}, not whether or not people act {\itshape per se}.'' \citep[p.~125]{brugger_proximising_2016}. They also explain that because values are considered high-level features, people with altruistic and biospheric values are more likely to participate in mitigation and adaptation actions if climate change is considered distant rather than proximal \cite{brugger_psychological_2015}.

Brügger et al., also propose two (opposing) reasons for why some studies do not show greater mitigation and adaptation efforts when psychological distance is reduced \cite{brugger_psychological_2015}. The first reason is that people’s actions depend on their feelings of place attachment and they would act regardless if that place was proximal or distant. ``That is, the more one is attached to a specific proximal place as a whole…, the more likely one is to become concerned about and respond to a message that conveys a threat to these cherished things. By contrast, people who do not relate in any way to such a place will most probably remain unaffected by proximized messages.'' \citep[p.~1033]{brugger_psychological_2015}. The opposing reason proposed by Brügger et al. is that proximizing threats about a place to which a person holds strong attachments can invoke fear which causes them to feel overwhelmed and therefore leads to defensive actions but not those that would help act on the perceived threat \cite{brugger_psychological_2015}. Instead of CLT, Brügger suggests alternative frameworks that can be used to study psychological distance related to climate change such as risk processing models, mental models, bayesian updating, and conceptual change \cite{brugger_understanding_2020}. 

Another facet to the mixed results seen in these studies is due to the varying methods to study how psychological distance impacts ``perceptions of climate change'', ``climate action'', and ``motivation'', ``intention'', and ``willingness''. For example, the perception of climate change can be studied by measuring whether individuals themselves will be affected by climate change as compared to how other people will be affected. Duan et al. find that more abstract images of climate change lead people to perceive climate change as more spatially distant to themselves but more proximal to others, e.g., citizens of the Maldives \cite{duan_refining_2022}. Especially when psychological distance to climate change is framed in egocentric terms (i.e., in relation to the self), low-level features lead to low-level construals which makes climate change more proximal than when framed in non-egocentric terms (i.e., in relation to others by putting yourself in the perspective of another) \cite{duan_refining_2022}.

Impacts of psychological distance to climate action have been examined by studying ``mitigation behavioural intentions'' which are climate change friendly behaviours such as carpooling, buying energy efficient appliances, using less air conditioning, and so on \cite{duan_refining_2022}. However, they can also be studied as ``willingness to take action'' through questions like ``willingness to make a donation to address climate change'' \citep[p.~147]{manning_psychological_2018}. Studies also measure ``willingness to sacrifice'' by asking participants questions about ``...willing to give up traveling in order to lower energy consumption…'', ``willing to pay a higher price for a product that is environmentally-friendly…'' etc. \cite{carmi_further_2015}. Brügger et al. studied adaptation intentions after presenting participants with near versus far framings of climate change such as ``buy a flood insurance for your (future) home'', ``donate money to preserve species at risk from climate change'', ``read about how to avoid heat stress during heat waves'', etc. \cite{brugger_proximising_2016}. 

While each of these studies fill gaps in research on the psychological distance to climate change, the lack of systematic organization of the concepts and concerns in the field lead to results that amplify disagreements rather than come to a unified roadmap of where future discourse should be focussed. 

Particularly, it is important to emphasize that individual behaviour is embedded in situations and structures that often act as barriers to action. As Steg points out, ``Behavior is influenced not only by motivational factors, such as values, environmental and climate concerns, and personal norms, but also by structural factors such as the availability of technologies, products, and infrastructures; price regimes; institutions; and laws and regulations. Such structural factors define the costs and benefits of choice options, which can have major implications for how attractive and feasible it is to engage in climate actions.'' \citep[p.~403]{steg_psychology_2023}. While psychological distance can help make information used to make decisions more proximal, this proximity would not guarantee `perfect' decisions. We should be aware of such techno-solutionist framings as those perpetuated by persuasive design that could be understood as manipulative. 
% it cannot directly influence decisions which are situated in social contexts. 

\vspace{-0.2cm}

\subsection{From Individual Self to Ecological Self} \label{indself_ecolself}

If we shift our view of the self from individualism to one that is a part of the environment or one that is embedded in our ecology, psychological distance may disappear. In {\itshape Active Hope} \citep[p.~115]{macy_active_2022}, Macy and Johnstone contrast ways of viewing the self -- the self could be a discrete entity with a clear outer boundary to the rest of the world, and the self could also be one that emerges from our relationships, contexts, communities interconnected with the web of life. They argue that the first way of viewing the self is harmful to personal well-being, community well-being, and planetary well-being, and that by embodying a larger story of who we are, we will be able to heal our world and communities \citep[p.~119]{macy_active_2022}. More specifically, the idea of the {\itshape ecological self}, first introduced by Arne Naess \cite{naess_self-realization_1987} to describe a sense of the self that includes the natural world, could provide a greater source of strength for supporting the well-being of our world because we are able to recognize a larger story of who we are and a longer time span of where we come from \citep[p.~121]{macy_active_2022}.

Although anthropocentrism and consumerism are still prevalent in the field of computing, a growing body of research sheds light on alternative paradigms such as entanglement HCI \cite{frauenberger_entanglement_2020}, more-than-human perspectives \cite{clarke_more-than-human_2018,clarke_more-than-human_2019}, and ecofeminist design \cite{churchill_feminist_2023}, which help us understand a broadening version of the self. Frauenberger argues for a new ontological approach to HCI that engages with entanglement theories and through them understands the inseparable nature of humans and things in their surroundings \cite{frauenberger_entanglement_2020}. Clarke et al. propose an alternative smart city agenda that decenters human agencies, explores temporalities of the more-than-human, incorporates other wisdom about the more-than-human, and includes design pedagogy and learning to mediate the more-than-human  \cite{clarke_more-than-human_2018,clarke_more-than-human_2019}. Stead et al. explored design considerations for more-than-human data interaction that refocuses the needs of ecological actants \cite{stead_more-than-human-data_2022}. Ecofeminist HCI scholars argue for critical examinations of ``how ecological responsibility is geopolitically negotiated, discursively operationalized and ethically justified based on specific values and priorities'' \citep[p.~22]{kannabiran_preamble_2023}. Feminist ecologies is both a design space and a discursive space to rethink the past and to imagine alternative futures, attending to nonhuman relations and designing with them to trouble oppressive systems \cite{churchill_feminist_2023, woytuk_menstrual_2023}.

Following arguments in anthropology, planning, geography and design, computing research is challenging views of design practices in computing as objective and neutral, looking beyond conventional binaries such as human/non-human and culture/nature, to understand computing’s roles and impact in the broader entanglements of all living beings and objects. To help move towards the ecological self, we again see speculative design as an avenue for exploration. Frauenberger relied on a speculated technology, Flow, to illustrate his paradigms on entanglement HCI and called for ``agonistic, participatory speculation methods to design meaningful relations, rather than optimizing user experiences'' \citep[p.~19]{frauenberger_entanglement_2020}. Clarke et al. proposed a participatory speculative urban walk to create an embodied experience for pushing forward a cultural change in smart cities \cite{clarke_more-than-human_2018}. Woytuk and Søndergaard also pointed out the importance of encouraging speculations around neglected or unknown relations \cite{woytuk_menstrual_2023}. Through envisioning alternative futures of computing beyond the current one dominated by ideas such as consumerism, capitalism and techno-solutionism, we see hopes and actions for working towards the flourishing of plural worlds with diverse forms of policymaking, knowing, and living with ecologies.

 \vspace{-0.2cm}

\section{Conclusion}

So how can psychological distance inform and shape data visualization design for policymakers? It would be a mistake to assume that distant effects simply do not matter to us – and a costly mistake since it leads to defeatism. Distant effects are neither irrelevant to us humans, nor do we benefit collectively from deceiving ourselves making them appear proximal. Psychological distance does however affect how we perceive, negotiate, reason, and decide. The structure of social, temporal, spatial, and hypothetical distance to a decision-maker can guide designers in identifying how decision support can best present information with full consideration of the dynamics of psychological distance. Two creative avenues are highlighted in this paper in the context of decision making related to climate change. The opportunity to pursue visceral representations of data evidently influences what kind of construal takes place, but the effects of that shift depend on context and the outcomes of such shifts will need to be evaluated carefully. Speculative design opens up creative avenues for envisioning discontinuous change and alternative lifeworlds. It is  noteworthy that climate change is only one aspect of our planetary boundaries \cite{persson_outside_2022,richardson_earth_2023,rockstrom_planetary_2024} and we hope our discussions focusing on climate change could open up discussion around the other boundaries, which might play out on different time, social, and physical scales.

Evidence shows that human beings are quite ready to cooperate and collectively build long-term strategies that benefit future generations even when it comes at a cost to the present \cite{krznaric_good_2020}. A groundbreaking study concluded that ``[m]any citizens are ready to sacrifice for the greater good. We just need institutions that help them do so.'' \citep[p.~222]{hauser_cooperating_2014}. Experienced decision-makers are highly skilled in recognizing and dealing with complexity \cite{klein_sources_2017}, and they are often remarkably perceptive of the conditions of temporal and social distance as well as the constraints their decisions are subject to \cite{fagerholm_its_2023}. It makes sense to treat them not as faulty programs but to explore jointly how they might best reorganize their decision-making environment to support responsible policymaking in times of urgent change. Speculative designs of data visceralizations appear to be one promising avenue by which we might bring Limits to the fore and find ways to flourish within them.

\section*{Acknowledgements}
This research was partially supported by NSERC through RGPIN-2016-06640, the Canada Foundation for Innovation, and the Ontario Research Fund.

\bibliographystyle{ACM-Reference-Format}
\bibliography{PsychDist_LIMITS24}

\appendix

\section{Literature Review Method} \label{ap_litreview}

Due to the multi-disciplinary nature of the concepts we synthesize, we lay out our methodology for conducting the literature review. Our literature review was composed of literature within 3 groups: psychological distance and environmental psychology, speculative design, and data visceralization. 

The material on psychological distance was first compiled by reviewing Christoph Becker’s, one of the authors, existing literature database compiled from an intertemporal choice project \cite{Becker_Walker_McCord_2017}, a second project on intertemporal choice in software engineering \cite{fagerholm_its_2023}, and a book chapter on intertemporal choices in sustainability \cite{christoph_becker_searching_2023}. This literature database was also informed by conversations and recommendations from behavioural economics researchers. In addition to this database, key starting references included The Wiley Blackwell Handbook of Judgment and Decision Making \cite{Keren_Wu_2015}, foundational papers on psychological distance and CLT, and author searches for works by Yaacov Trope and Nira Liberman. The other authors found additional starting references through Annual Review of Psychology papers on ‘Psychology of Climate Change’ \cite{steg_psychology_2023} and environmental psychology \cite{gifford_environmental_2014}. 

Initial sources for speculative design literature came from class readings from a speculative design course at University of Toronto and Soden et al.’s paper \cite{soden_what_2021}. 

Initial sources for literature on data visceralization included the books The Visual Display of Quantitative Information \cite{tufte_visual_2013} and Data Feminism \cite{dignazio_rational_2020} from which the term `data visceralization' was made popular. Additional literature was added to the list through discussions with visualization researchers over the past year. The authors of this paper were exposed to various case studies leveraging VR, physicalization and techniques of visual representation to create more effective data visualizations that speak to affect and emotions.

The second stage for collecting literature consisted of Google Scholar searches using search terms including:
\begin{itemize}
    \item Psychological distance, environmental psychology
    \item Data visceralization, visceral data, affective design
    \item Speculative design, speculation, design fiction
\end{itemize}

The final stage included identifying additional literature through forward and backward snowballing from previously collected sources.

\end{document}